\documentclass[preprintnumbers,aps,prb,preprint,showpacs,showkeys]{revtex4}

\usepackage{graphicx}
\usepackage{amsmath}
\usepackage{amssymb}
\usepackage{amsfonts}

\begin{document}

\preprint{APS/123-QED}

\title{Analytical Approach to the Local Contact Potential Difference on (001) Ionic Surfaces:~Implications for Kelvin Probe Force
Microscopy}

\author{Franck Bocquet$^{1, 2}$, Laurent Nony$^{1, 2, }$\footnote{To whom correspondence should be addressed; E-mail:
laurent.nony@im2np.fr.}, and Christian Loppacher$^{1, 2}$}
\affiliation{$^{(1)}$Aix-Marseille Universit\'{e}, IM2NP, Centre
Scientifique de Saint-J\'{e}r\^{o}me, Avenue Escadrille
Normandie-Niemen, Case 151, F-13397 Marseille CEDEX 20, France\\
$^{(2)}$CNRS, IM2NP (UMR 6242), F-13397 Marseille-Toulon, France}

\author{Thilo Glatzel}
\affiliation{Department of Physics, University of Basel,
Klingelbergstr.~82, CH-4056 Basel, Switzerland}

\date{\today}

\pacs{07.79.Lh, 41.20.Cv, 73.40.Cg}

\keywords{non-contact Atomic Force Microscopy; Kelvin Probe Force
Microscopy; Local Contact Potential Difference; Short-range
Electrostatic Force; Analytical model; Madelung Surface Potential;
Ionic crystal}

Published in Phys. Rev. B \textbf{78}, 035410 (2008)

\begin{abstract}
An analytical model of the electrostatic force between the tip of
a non-contact Atomic Force Microscope (nc-AFM) and the (001)
surface of an ionic crystal is reported. The model is able to
account for the atomic contrast of the local contact potential
difference (CPD) observed while nc-AFM-based Kelvin Probe Force
Microscopy (KPFM) experiments. With the goal in mind to put in
evidence this short-range electrostatic force, the Madelung
potential arising at the surface of the ionic crystal is primarily
derived. The expression of the force which is deduced can be split
into two major contributions: the first stands for the coupling
between the microscopic structure of the tip apex and the
capacitor formed between the tip, the ionic crystal and the
counter-electrode; the second term depicts the influence of the
Madelung surface potential on the mesoscopic part of the tip,
independently from its microscopic structure. The former has the
lateral periodicity of the Madelung surface potential whereas the
latter only acts as a static component, which shifts the total
force. These short-range electrostatic forces are in the range of
ten pico-Newtons. Beyond the dielectric properties of the crystal,
a major effect which is responsible for the atomic contrast of the
KPFM signal is the ionic polarization of the sample due to the
influence of the tip/counter-electrode capacitor. When explicitly
considering the crystal polarization, an analytical expression of
the bias voltage to be applied on the tip to compensate for the
local CPD, \emph{i.e.} to cancel the short-range electrostatic
force, is derived. The compensated CPD has the lateral periodicity
of the Madelung surface potential. However, the strong dependence
on the tip geometry, the applied modulation voltage as well as the
tip-sample distance, which can even lead to an overestimation of
the real surface potential, makes quantitative KPFM measurements
of the local CPD extremely difficult.
\end{abstract}

\maketitle
\section{Introduction}
Electrostatic forces play a key role in non-contact Atomic Force
Microscopy (nc-AFM), not only in the imaging
process~\cite{sadewasser03a} but also for the investigation of the
surface electronic properties. Electronic properties such as the
work function and surface charges can be acquired by Kelvin Probe
Force Microscopy (KPFM)~\cite{weaver91a,nonnenmacher91a}
simultaneously to imaging topography by nc-AFM. In KPFM, a
feedback is used to apply a voltage between the tip and the sample
in order to minimize the electrostatic tip-sample interaction. For
metals, this voltage is equal to the contact potential difference
(CPD), \emph{i.e.} the bias voltage to be applied between the tip
and the surface to align their fermi levels. It is connected to
the difference between the work functions of the two surfaces, and
thereby to their local electronic properties, according to:

\begin{equation}\label{EQU_LOCAL_CPD}
\Delta \phi=\phi_{tip}-\phi_{sample}=qV_{cpd},
\end{equation}

\noindent $q$ being the elementary electrical charge: $q=1.6
\times 10^{-19}$~C.

Nowadays, two KPFM-based techniques provide facilities to map the
spatial variations of the CPD on the nanometer scale, namely
Frequency-~\cite{kitamura98a} or
Amplitude-Modulation-KPFM~\cite{kikukawa96a, sommerhalter99a,
glatzel03a} (FM- or AM-KPFM, respectively). These methods were
primarily applied to metallic and semiconducting surfaces to study
the distribution of dopants in semiconductors \cite{kikukawa95a},
or the adsorption of organic molecules (for an overview see
ref.[\onlinecite{palermo06a}]). In a few experiments, even
molecular~\cite{sasahara01a} or
atomic~\cite{sugawara99a,kitamura00a,okamoto03a} contrast has been
reported. The extension of the technique to insulating surfaces,
was performed more recently, as demonstrated by the results
reported on thin ionic films on metals
\cite{krok04a,loppacher04a}, or on the contribution of bulk
defects to the surface charge state of ionic crystals
\cite{barth06a,barth07a}.

\begin{figure}[t]
  \includegraphics[width=\columnwidth, angle=0]{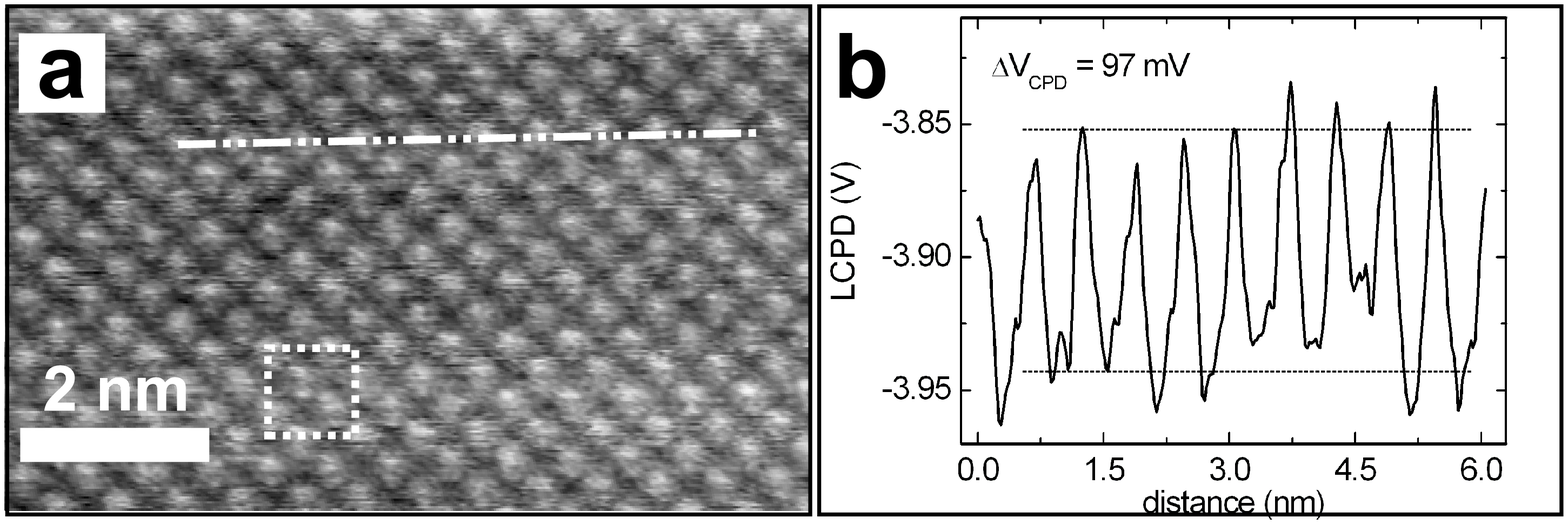}\\
  \caption{a- Experimental image showing the atomic contrast of the compensated CPD measured on a (001) surface of KBr in ultra-high vacuum by means of AM-KPFM. The vertical contrast ranges from -3.95 to -3.85~V from dark to
white
  spots. The dashed line depicts the cross section shown in b-. The dotted square depicts the area corresponding to the ball model shown in fig.\ref{FIG_GEOM}a. b- Corresponding cross section.}\label{FIG_EXP}
\end{figure}

In this work, atomic corrugation of the CPD signal is reported for
the first time on the (001) surface of a bulk ionic crystal of
KBr. For that purpose, KPFM experiments were carried out in
ultrahigh vacuum with a base pressure below $10^{-10}$~mbar using
a home built non-contact atomic force microscope operated at room
temperature \cite{howald93a}. A highly doped silicon cantilever
with a resonance frequency $f_0\approx 160$~kHz, and a spring
constant $k\approx 21$~N.m$^{-1}$ was used. The typical
oscillation amplitude of the fundamental bending resonance was
$\approx 10$~nm. The cantilever was annealed
(30~min~@~120$^\circ$) and gently sputtered with Ar$^+$ ions
(1-2~min~@~680~eV). The KBr crystal was cleaved in ultrahigh
vacuum along the (001)-plane and subsequently annealed at
120$^\circ$ during an hour. The KPFM signal was detected using the
AM mode, as described in detail in Ref. \onlinecite{kikukawa96a}.
While these experiments, atomic-scale contrast was as well visible
in the topography channel (data not shown). The CPD measurements
are reported in figs.\ref{FIG_EXP}a and b. In fig.\ref{FIG_EXP}a,
the image exhibits atomic features, the measured period of which
is 0.63~nm, which is visible in the joint cross section. This
value matches to a good agreement the lattice constant of KBr,
0.66~nm. The vertical contrast yields about 100~mV around an
average value of -3.9~V, the origin of which will be discussed in
section \ref{SEC_KPFM}. A striking aspect of those results is the
robustness of the imaging process in terms of stability and
reproducibility upon the tips used. These results suggest an
intrinsic imaging process relying on the microscopic origin of the
contact potential arising at the sample surface. In this case, the
CPD rather turns into \emph{local CPD}, consistently with the
concept of \emph{local work function} which has been introduced by
Wandelt on metals\cite{wandelt97a}. By ``intrinsic imaging
process", it is meant that the contrast can be accounted for with
a tip consisting of a single material, namely a metal. Thus, the
atomic contrast neither relies on adsorbed nor on unstable species
at the tip apex, as often reported for
topographic\cite{hoffmann04a} or dissipation\cite{hoffmann07a}
imaging by nc-AFM.

In this work, in order to understand the local CPD contrast
formation, an electrostatic model is proposed that allows us to
derive an expression of the short-range electrostatic force
occurring between a biased metallic tip of a nc-AFM microscope and
the surface of a bulk dielectric. On the contrary to more refined
numerical methods proposed in the literature for almost similar
systems\cite{kantorovich_SciFi,lyuksyutov04a}, the analytical
development is restricted to a simple tip geometry and a
classical, continuous electrostatic approach. Notwithstanding, the
model allows us to define a general frame, able to put in relation
the surface electrostatic properties with the imaging process
yielding atomic contrast of the CPD on ionic surfaces. Obviously,
the main results presented here remain qualitatively correct for
more complex tip geometries, although numerical methods are then
required to get quantitative numbers.

The motivations for that work are twofold. On the one hand, to our
knowledge, a compact modelization of the short-range electrostatic
forces responsible for the atomic contrast in KPFM on a bulk ionic
crystal is still lacking. On the other hand, when evaluating KPFM
experiments, the relative complexity of FM- or AM-KPFM
experimental setups, both including four electronic controllers,
makes the interpretation of the experimental images, and primarily
CPD images, tedious, especially when dealing with atomic
resolution. Several groups analyzed the KPFM imaging mechanism in
order to evaluate their data in terms of quantitative values and
lateral resolution\cite{jacobs98a, colchero01a, mcmurray02a,
sadewasser03b, rosenwaks04a, takahashi04a, palacios05a,
zerweck05a, leendertz06a, zerweck07a}, a few of them also compared
AM- and FM-KPFM in terms of evaluating the force and its gradient,
respectively\cite{glatzel03a, zerweck05a}. The fact that many of
the above mentioned calculations point out that KPFM results,
especially for nano-objects, show a strong distance dependence,
clearly points out that for a careful analysis, it is not
sufficient to only calculate the electrostatic tip-sample
interaction. It is rather important to perform simulations
including all imaging mechanisms and in particular also the
distance control in order to exclude artifacts due to the feedback
circuits. In order to explore the origin of the CPD atomic
contrast, we are aiming to closely mimic a real KPFM setup by
means of an earlier developed nc-AFM simulator\cite{nony06a}. In
the present case of an ionic surface, it is required to consider a
large slab of ions, virtually infinite, to describe properly the
electrostatic interaction. This is hardly feasible by means of
\emph{ab initio} calculations which fail to describe tip-surface
systems involving a too big number of atoms. Therefore, prior to
simulating the CPD contrast on ionic surfaces by means of our
simulator, which is the scope of a future work, we have to find an
analytical expression for the electrostatic tip-sample
interaction.

The following section details the boundary-value electrostatic
problem leading to the expression of the force (section
\ref{SEC_FORCE}). In section \ref{SEC_DISCUSSION}, the analytical
expression of the local CPD is derived and discussed to be put in
relation with the experimental observations. The implications for
KPFM experiments are discussed as well.

\section{Electrostatic model}\label{SEC_MODEL}

\begin{figure}[t]
  \includegraphics[width=\columnwidth, angle=0]{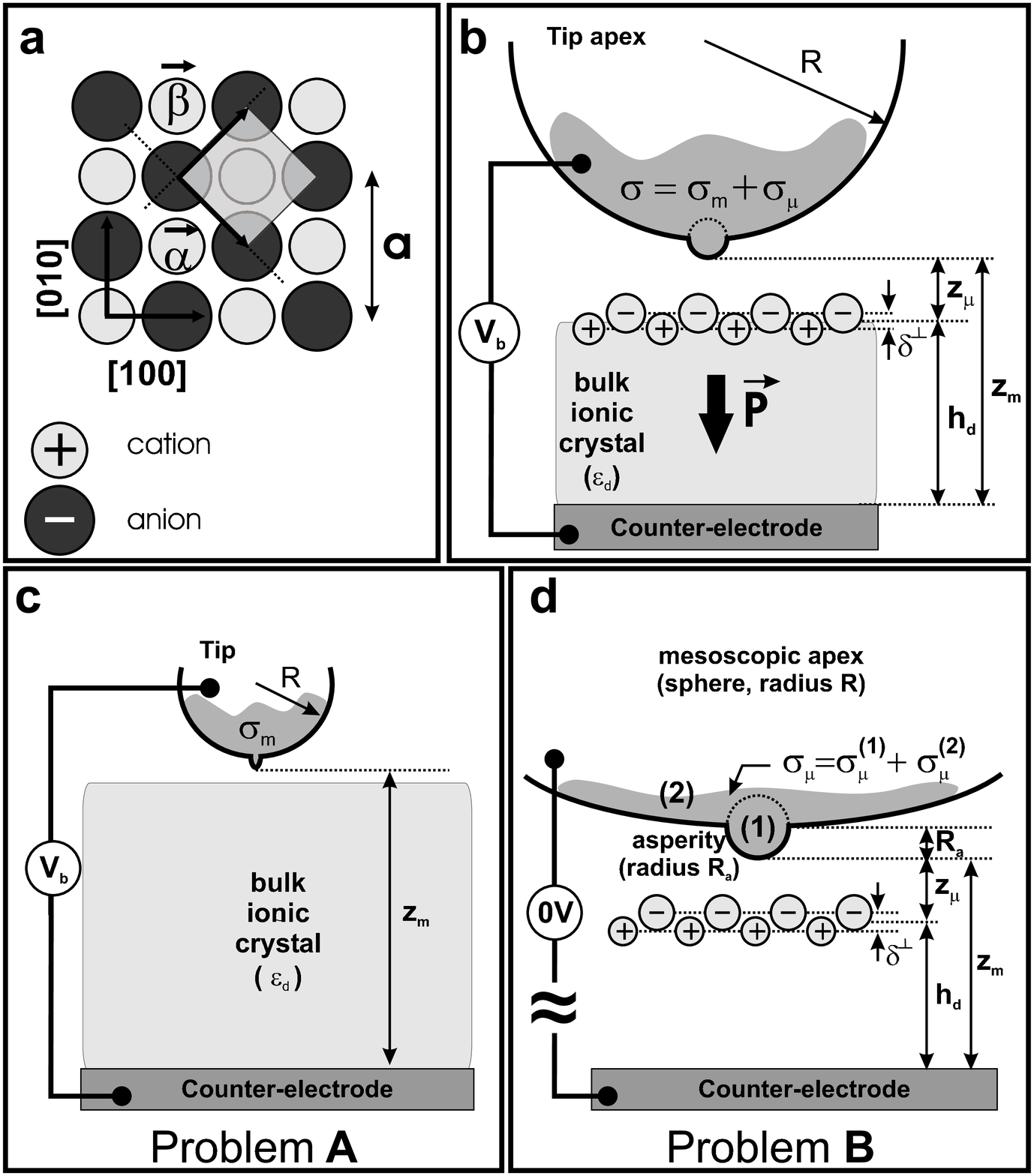}\\
  \caption{a- KBr lattice with a fcc structure corresponding to
  the dotted square shown in fig.\ref{FIG_EXP}a. The white spots of the experimental image have been placed on top of anions arbitrarily. b- Scheme of the KPFM experiment defining the electrostatic boundary-value problem to be solved. The metallic tip is biased with respect to a metallic counter-electrode placed a few millimeters
  far from it owing to the thickness of the ionic crystal. The bias voltage polarizes the crystal, which results, at the surface, in a modulation of the
  positions of the ions. c- and d- Schemes of the splitting of the original electrostatic boundary-value problem schemed in b- defining problems A and B, respectively.}\label{FIG_GEOM}
\end{figure}

The geometry of the problem of FM- or AM-KPFM experiments applied
to bulk insulating materials is reported in fig.\ref{FIG_GEOM}b.
The dielectric sample is an alkali halide crystal like NaCl, KBr,
KCl with a fcc structure and a lattice constant $a$ (\emph{cf.}
fig.\ref{FIG_GEOM}a). In the area where the tip is, the model
assumes that the crystal carries neither net charge, nor local
dipole. Its thickness $h_d$ is much larger than all other
distances of the problem, typically a few millimeters. Below the
surface, the crystal is treated as a continuous dielectric medium
with a dielectric permittivity $\epsilon_d$. At the surface, the
atomic corrugation of the crystal is described by a single layer
of alternate point charges arranged with a fcc structure
perpendicular to the [001] direction. The layer extends infinitely
in the plane direction. The motivation for such a rationalization
of the problem will be justified in section (\ref{SEC_SIGMA_MU}).

The crystal lies on a metallic sample holder (hereafter referred
to as the counter-electrode), with respect to which the tip is
biased. The counter-electrode is a planar and perfect conductor.
The tip is also assumed to be a perfect conductor which is biased
at $V_b$. In order to preserve the analyticity of the model, the
electrostatic boundary-value problem is restricted to a tip with a
very simple apex geometry, namely: a hemispherical mesoscopic part
(radius $R \simeq 5$~nm) on top of which is half-embedded a small
spherical asperity (radius $R_a \ll R$). The contributions of the
cantilever and of the macroscopic part of the tip to which the
apex is connected to are assumed to be negligible. This issue will
be justified in section \ref{SEC_FORCE_M}. The tip-surface
distance is denoted $z_\mu$ (typically a few~\AA).

The electric field $\overrightarrow{E}$ produced locally between
the tip, the dielectric, and the counter-electrode polarizes the
ionic crystal, which acquires a macroscopic polarization
$\overrightarrow{P}=n_v\overrightarrow{p_l}$ oriented along
$\overrightarrow{E}$. In the former equation, $n_v$ is the number
of polarizable species \emph{per} unit volume and
$\overrightarrow{p_l}$, the local dipolar moment \emph{per }unit
cell. In KPFM, the bias voltage $V_b$ is modulated at frequencies
which do not influence the electronic polarization of the ions.
The major part of the crystal polarization has rather an ionic
character, \emph{i.e.} a net displacement in opposite directions
of the ions due to their charge $\pm q$ with respect to their
equilibrium positions, $\pm \delta/2$, with $\delta = p_l/q$, $q$
being the elementary electrical charge. The polarization effect
occurs as well at the crystal surface, where the positions of the
ions become modulated perpendicularly to the surface plane,
\emph{i.e.} $\delta \rightarrow \delta^\bot$, as sketched in
figs.\ref{FIG_GEOM}b and d. $\delta^\bot$ is proportional to the
local electric field $\overrightarrow{E_l}$ at any ionic site and
to the total polarizability of the dielectric restricted, in our
approach, to the ionic polarizability $\alpha_i$
(ref.[\onlinecite{Note_Nony07b}]). Usually, $\overrightarrow{E_l}$
differs from the external electric field $\overrightarrow{E}$ due
to the biased tip because $\overrightarrow{E_l}$ explicitly
depends on the polarization of the dielectric. The Lorentz's model
links $\overrightarrow{E_l}$ and $\overrightarrow{E}$
(ref.[\onlinecite{kittel}]), thus, $\overrightarrow{p_l}$ is
written:

\begin{equation}\label{EQU_POLARIZATION}
\overrightarrow{p_l}=\alpha_i\epsilon_0
\overrightarrow{E_l}=\chi_d\overrightarrow{E}=q\delta^\bot
\frac{\overrightarrow{E}}{E},
\end{equation}

\noindent with $\chi_d=\alpha_i\epsilon_0/(1-n_v\alpha_i/3)$, the
dielectric susceptibility of the sample. In the former equation,
it is important to notice that $\overrightarrow{p_l}$ depends on
$\overrightarrow{E}$ and hence on the bias voltage $V_b$.
Consequently, this is also true for $\delta^\bot$. For the sake of
discussions, the bias dependence will henceforth be explicitly
outlined $\delta^\bot \rightarrow \delta^\bot(V_b)$. Despite
$\delta^\bot(V_b)$ cannot be estimated at this point, it is
crucial to keep in mind that the sample surface is polarized by
the influence of the bias since this is a key aspect of the origin
of the CPD atomic contrast.

Our approach of the electrostatic boundary-value problem relies on
an \emph{ad hoc} assumption. The tip, being a perfect conductor,
develops a surface charge density $\sigma$, the origin of which is
split into two main contributions $\sigma=\sigma_m+\sigma_\mu$,
namely:

\begin{itemize}
    \item a charge density $\sigma_m$ due to
the capacitor formed between the biased tip and the
counter-electrode with the dielectric in between (\emph{cf.}
fig.\ref{FIG_GEOM}c). Owing to the distance between the
electrodes, \emph{i.e.} roughly the dielectric thickness $h_d$,
$\sigma_m$ has a mesoscopic character. It is not influenced by the
local structure of the tip apex, but rather by its overall shape.
    \item a charge density $\sigma_{\mu}$ originating from the
    Madelung potential that expands at the crystal surface. When the tip is at a distance $z_\mu$ where the potential is effective, typically a few
    \AA ngstr\"{o}ms, then it develops, in addition to $\sigma_m$, a surface charge density $\sigma_\mu$ (\emph{cf.} fig.\ref{FIG_GEOM}d). $\sigma_{\mu}$ has a microscopic
    character and must strongly depend on the local structure of the tip apex and on $z_\mu$.
\end{itemize}

Despite the simple tip geometry that has been assumed, the
calculation of the electrostatic force acting on it due to the
combined influence of the capacitive coupling and of the
 Madelung surface potential has no exact analytical solution.
Nevertheless, one can build up an approximate solution to the
boundary-value problem relying on the superposition principle. For
that purpose, the problem is split up into two boundary-value
sub-problems schemed in figs.\ref{FIG_GEOM}c and d:

\begin{itemize}
    \item problem $A$: the tip biased at $V_b$ in
front of a dielectric continuous medium (height $h_d$, dielectric
permittivity $\epsilon_d$) held on an infinite planar conductor,
the counter-electrode (fig.\ref{FIG_GEOM}c). The local structure
of the dielectric is not supposed to influence the tip. This is
the description of the ``capacitive" problem, the solution of
which provides the surface charge density $\sigma_m$.
    \item problem $B$: the tip now biased at 0~V with an infinite plane
of alternate point charges located at the same distance than the
surface of the dielectric in problem A, \emph{i.e.} $z_\mu$. The
layer of point charges is polarized under the electric field that
occurs in problem A (fig.\ref{FIG_GEOM}d). This describes the
``microscopic" problem. The solution provides the surface charge
density $\sigma_\mu$.
\end{itemize}

Besides, in order to carry out the calculations more easily, it is
convenient to distinguish the two following geometrical areas on
the tip (\emph{cf.} fig.\ref{FIG_GEOM}d), namely: the asperity,
area (1), \emph{i.e.} a half-sphere with radius $R_a$ and the
mesoscopic tip apex around it, area (2), a sphere with radius $R
\gg R_a$ (typically $R/R_a \geq 50$). These two areas do not
overlap, but the continuity between them is ensured. The vertical
force acting on the tip\cite{jackson} is written in terms of
$\sigma_m$ and $\sigma_\mu$ as described above:

\begin{equation}
F=\int_\text{tip}\frac{\left(\sigma_m+\sigma_{\mu}
\right)^2}{2\epsilon_0}\widehat{n}.\widehat{u}_zd\mathcal{S} =
F_{m}+F_{m\mu}+F_{\mu}
\end{equation}

\noindent $\widehat{n}$ and $\widehat{u}_z$ are the normal to the
tip surface and the unitary vector along the vertical $z$ axis of
the problem, respectively. Doing so, we only focus on the vertical
resultant of the force acting onto the tip. The above expression
can be expanded into three components: a purely capacitive part,
$F_{m}$ originating from the tip/dielectric/counter-electrode
capacitor; a coupling part, $F_{m\mu}$, which can be interpreted
in terms of the resulting force of all the elementary forces due
to the electric field $\sigma_m/\epsilon_0$ onto each elementary
charge $\sigma_{\mu} d\mathcal{S}$ produced on the tip by the
influence of the Madelung potential of the crystal, $V_s$; and a
purely microscopic part, $F_\mu$, standing for the short-range
electrostatic force due to $V_s$.

\subsection{Problem A: estimation of $\sigma_m$}\label{SEC_PROBLEM_A}
Although the boundary-value problem of a planar conductor biased
with respect to another one with an incomplete dielectric layer in
between yields an expression of the surface charge density, the
problem with the sphere does not. However, one can argue that the
expression of $\sigma_m$ must be a combination between a
configuration in which there is no dielectric medium in the
sphere/counter-electrode interface and an opposite one, where the
interface is completely filled with it. One can therefore
postulate an effective dielectric permittivity
$\widetilde{\epsilon}_d=K\epsilon_d$, where $K~(<1)$ is a constant
to be set. Owing to the fact that the mesoscopic part of the tip
apex, referred to as area (2), is located at a distance
$z_m=z_\mu+h_d \gg R$ from the counter-electrode, the analytical
expression of $\sigma_m^{(2)}$, explicitly calculated in
refs.[\onlinecite{jackson, durand}], asymptotically trends towards
the surface charge density of an isolated, biased sphere
\cite{jackson}:

\begin{equation}\label{EQU_SIGMA_M2}
\sigma_m^{(2)} \overset{z_m \gg R}{=}
\frac{\widetilde{\epsilon}_d\epsilon_0 V_b}{R}
\end{equation}

\noindent To get the surface charge density on area (1),
$\sigma_m^{(1)}$, we seek the potential $V_m^{(1)}$ which equals
$V_b$ over the asperity. For that purpose, a spherical coordinate
system $(r,\theta,\varphi)$ centered on the asperity is chosen.
The problem having an azimuthal symmetry, the sought potential can
be expanded in Legendre polynomials \cite{jackson}:

\begin{equation}
V_m^{(1)}(r,\theta)=\sum_{n=0}^{\infty}\left( \alpha_n r^n +
\frac{\beta_n}{r^{n+1}} \right) P_n(\cos \theta)
\end{equation}

\noindent At large distance from the asperity, $r \gg R_a$, the
potential must be similar to the one of a sphere with radius $R$
biased at $V_b$, namely \cite{jackson}:

\begin{equation}
V_m^{(1)}(r,\theta) \overset{R>r\gg
R_a}{=}V_b\sum_{n=0}^{\infty}\left(\frac{r}{R}\right)^n P_n(\cos
\theta)
\end{equation}

\noindent Hence, the coefficients $\alpha_n$ of the expansion are
known. The coefficients $\beta_n$ are deduced from the property of
orthogonality of the Legendre polynomials at the boundary
condition $V(r=R_a)=V_b$. The potential is finally written:

\begin{equation}\label{EQU_POTENTIAL_Vm1}
V_m^{(1)}(r,\theta)=V_b\left\{ 1+ \sum_{n=1}^\infty
\left[\left(\frac{r}{R}\right)^n-\frac{R_a^{2n+1}}{R^n r^{n+1}}
\right]P_n(\cos \theta) \right\}
\end{equation}

\noindent The former equation rigorously describes the potential
of a system apex/asperity where the junction point between the two
spheres is smoothed and not singular, as sketched in
fig.\ref{FIG_GEOM}d. Nevertheless, for $R \gg R_a$, the influence
of the singular area is negligible. Therefore,
equ.\ref{EQU_POTENTIAL_Vm1} is a good approximation to the
boundary-value problem. The normal derivation along the surface of
the asperity yields the expression of $\sigma_m^{(1)}$:

\begin{equation}\label{EQU_SIGMA_M1}
\sigma_m^{(1)}= -\frac{\widetilde{\epsilon}_d\epsilon_0
V_b}{R}\sum_{n=1}^{\infty}(2n+1)\left(\frac{R_a}{R}\right)^{n-1}P_n(\cos
\theta)
\end{equation}

\noindent Owing to the condition $R \gg R_a$, the sum can be
restricted to the first term. Therefore:

\begin{equation}
\sigma_m^{(1)}= -\frac{3\widetilde{\epsilon}_d\epsilon_0
V_b}{R}\cos \theta
\end{equation}

\noindent Thus, despite $\sigma_m^{(1)}$ develops on the asperity,
its strength is governed by the radius of the mesoscopic part of
the tip apex, $R$ and not by the local radius of curvature of the
asperity, $R_a$.

\subsection{Problem B: estimation of $\sigma_{\mu}$}\label{SEC_SIGMA_MU}
The calculation of the surface charge density $\sigma_{\mu}$ on
the mesoscopic sphere+asperity is more tedious, primarily because
it relies on the estimation of the Madelung potential of the ionic
crystal, $V_s$. The boundary-value problem is now restricted to
the determination of the surface charge density arising on a
metallic tip at zero potential under the influence of an infinite
planar slab of point charges. Again, the solution of such a
problem has no straightforward analytical solution. However, we
can again use the condition $R \gg R_a$, as depicted, to some
extend, in fig.\ref{FIG_GEOM}d. Consequently, the mesoscopic part
of the apex can be assumed as equivalent to an infinite planar
conductor, at least in a small area along the sides of the
asperity (light grey area in fig.\ref{FIG_GEOM}d). The new
boundary-value problem defined by an infinite planar conductor
influencing another infinite planar conductor at zero potential
carrying a hemispherical bump with a radius $R_a$, now yields a
solution \cite{durand}. The method of the image charges is used to
solve it. The first set of image charges ensuring a zero-potential
value on the counter-electrode\cite{Note_Nony07f} produces an
electric field which influences the tip. But as a matter of fact,
this contribution can be neglected because the distance between
the tip and the counter-electrode is on the millimeter range and
the image charges originate from the Madelung potential, which is
known to decay exponentially fast\cite{watson81a, Note_Nony07c}
(\emph{cf.} also hereafter). The second set of image
charges\cite{Note_Nony07g} is quasi-punctual and located at the
center of the asperity. Hence, owing to the simplified geometry of
the electrode, the problem is reduced to a sphere with radius
$R_a$ at zero potential in the influence of two infinite planes,
\emph{i.e.} the slab and its image, which are anti-symmetrically
spaced with respect to the sphere. This procedure ensures a
zero-potential on the approximated plane within which the asperity
is embedded. Again, it is more convenient to use a spherical
coordinate system centered on the asperity. Then, the surface
charge density
$\sigma_{\mu}=\sigma_{\mu}^{(1)}+\sigma_{\mu}^{(2)}$ is derived by
normal derivation along areas (1) and (2), namely:

\begin{equation}\label{EQU_SIGMA_MU}
\sigma_{\mu}=-\left.\underbrace{\epsilon_0 \frac{\partial
V_\mu(r,\theta,\varphi)}{\partial
r}}_\text{(1)=asperity}\right|_{r=R_a}-\left.\underbrace{\epsilon_0\frac{\partial
V_\mu(r,\theta,\varphi)}{r\partial \theta}}_\text{(2)=planar
area}\right|_{\theta=\pi/2},
\end{equation}

\noindent The potential $V_\mu$ is derived from the Madelung
potential of the ionic crystal, previously referred to as $V_s$,
by the method of the image charges and the Kelvin transform
(influence on a sphere biased at zero potential)\cite{durand},
namely:

\begin{equation}\label{EQU_KELVIN_TRANSFORM}
\begin{array}{l}
V_\mu(r,\theta,\varphi)=\left.\left\{V_s(r,\theta,\varphi)-\frac{R_a}{r}V_s\left(\frac{R_a^2}{r},\theta,\varphi
\right)\right\}\right|_\text{slab}-\\
\left.\left\{V_s(r,\pi-\theta,\varphi)-\frac{R_a}{r}V_s\left(\frac{R_a^2}{r},\pi-\theta,\varphi
\right)\right\}\right|_\text{image slab}\end{array}
\end{equation}

\noindent The former equation fulfills the boundary condition
$V_\mu(r=R_a)=0$ everywhere along the surface asperity or along
the surface of the local planar area around it.

$V_s$ can be estimated on the base of the work by Watson \emph{et
al.} \cite{watson81a}. When considering an infinite planar slab of
point charges, the authors state that the potential, so-called
Madelung surface potential, reaches its asymptotic value in a very
short distance normal to the slab. Consequently, the ions within
the crystal at a distance only one lattice constant from the
surface have Madelung potentials which are indistinguishable from
those of the bulk. In other words, \emph{the tip will mainly be
influenced by the surface potential and not by the one arising
from the bulk part of the ionic crystal.} This is why a single,
infinite, layer of point charges is enough to describe the
influence of the Madelung surface potential on the tip, which
motivates our initial assumption. The potential is written:

\begin{equation}\label{EQU_MADELUNG_POT}
V_s\left(\overrightarrow{\rho},z_\mu\right)=\frac{1}{4\pi
\epsilon_0}\left(\frac{2\pi}{a'^2}\sum_{\overrightarrow{G}}q(
\overrightarrow{G})e^{i\overrightarrow{G}.\overrightarrow{\rho}}e^{-Gz_\mu}
\right),
\end{equation}

\noindent where $\overrightarrow{\rho}=x\widehat{i}+y\widehat{j}$
is the polar vector of any ion of the surface slab in an
orthogonal basis $(O,\widehat{i},\widehat{j})$, $O$ being the
projection of the center of the asperity on the surface,
$\widehat{i}$ and $\widehat{j}$ the unitary vectors of the fcc
unit cell. The summation is performed over the vectors
$\overrightarrow{G}$ of the reciprocal lattice of an arbitrarily
defined unit cell and $a'$ is proportional to the lattice constant
$a$ of the fcc unit cell. $q(\overrightarrow{G})$ is a structure
factor:

\begin{equation}\label{EQU_STRUCTURE_FACTOR}
q(\overrightarrow{G})=\frac{1}{G}\sum_k q_k
e^{i\overrightarrow{G}.\overrightarrow{\delta}_k^\|}e^{G
\delta_k^\bot}
\end{equation}

\begin{figure}[t]
 \includegraphics[width=\columnwidth, angle=0]{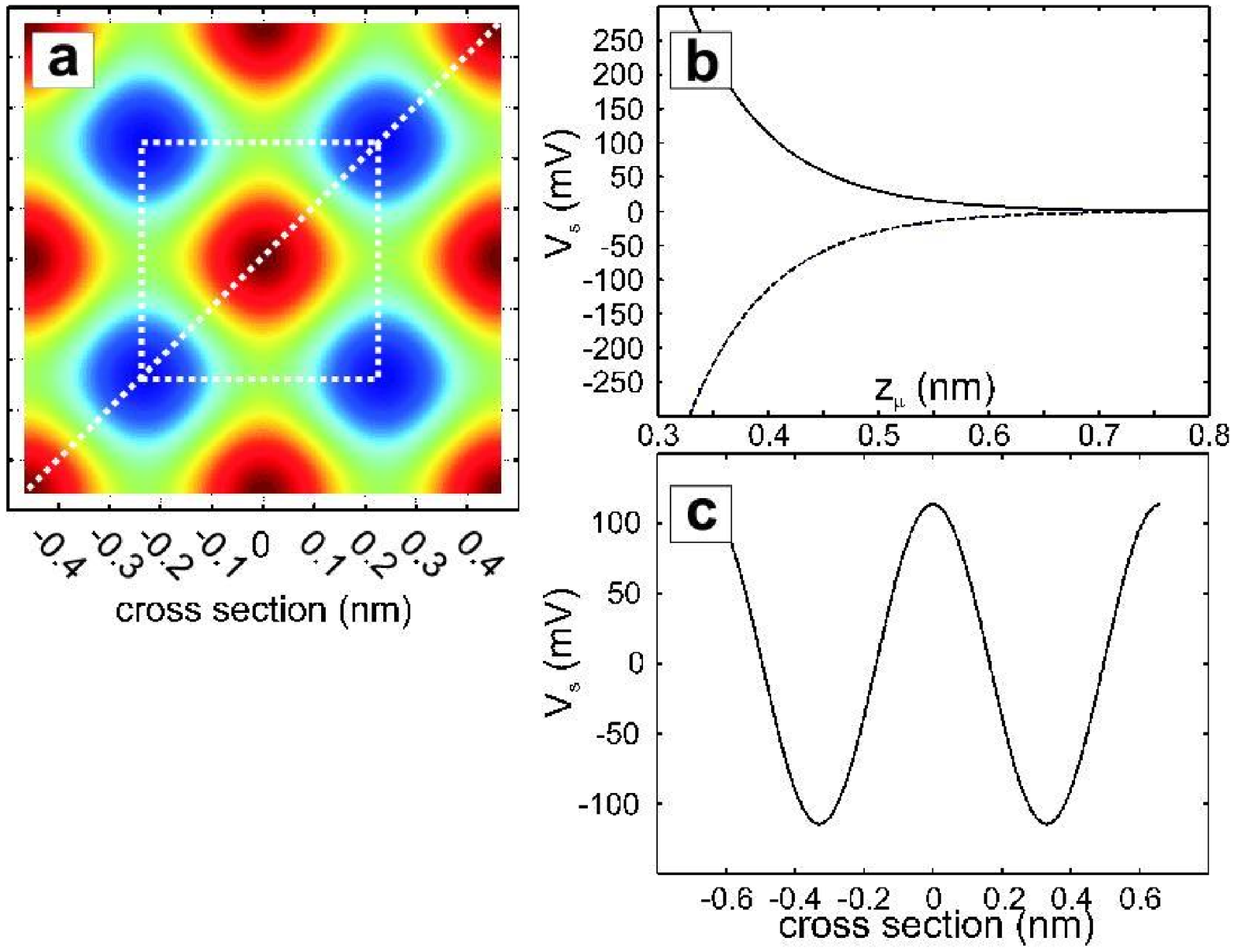}\\
 \caption{(Color online). a- Madelung surface potential calculated from equ.\ref{EQU_MADELUNG_POT_XPLICIT} for $a=0.66$~nm, $\delta^\bot=11$~pm and $z_\mu=4$~\AA.
 The vertical contrast ranges from $-100$ (blued spots) to $+100$~mV (reddish spots). The unit cell depicted with a dotted line is centered on a cation.
 b- Distance dependence of the potential on top of an anion (dotted curve) and on top of a cation (continuous curve) showing the exponential decay of the potential.
 c- Cross section along the dotted line shown in a-.}\label{FIG_POT}
\end{figure}

\noindent It is summed over the ions within the defined unit cell.
The k$^{th}$ ion carries an electrical charge $q_k$. Its planar
and perpendicular coordinates from the origin of the cell are
given by the two vectors $\overrightarrow{\delta}_k^\|$ and
$\overrightarrow{\delta}_k^\bot$. The latter reflects the
polarization effect felt by the ion within the unit cell,
previously referred to as $\delta^\bot(V_b)$. In
equ.\ref{EQU_STRUCTURE_FACTOR}, it is assumed that the
polarization of the ions extends all over the (001) surface plane.
In any case, it must extend over an much larger area than the tip
asperity. This assumption is consistent with the electric field
produced by area (2), which is constant within an area roughly
scaling as the mesoscopic tip radius $R$. $V_s$ is calculated from
the unit cell defined in fig.\ref{FIG_GEOM}a (light grey). It
consists of 4 anions and a cation weighting for a fourth and one,
respectively. The vectors of the direct lattice are
$\overrightarrow{\alpha}=a'\widehat{i}$ and
$\overrightarrow{\beta}=a'\widehat{j}$, where $a'=a\sqrt{2}/2$.
Owing to the exponential decay of the potential
 with $z_\mu$, visible in equ.\ref{EQU_MADELUNG_POT}, the calculation of the structure factor can be restricted to the first
four reciprocal vectors, namely: $\overrightarrow{G}_{i \text{ or
} j}^{\pm}=\pm 2\pi /a'(\widehat{i} \text{ or } \widehat{j}$). The
calculation yields:

\begin{equation}\label{EQU_MADELUNG_POT_XPLICIT}
V_s(x,y,z_\mu)= -\frac{q}{\pi
\epsilon_0a'}\cosh[\widetilde{\delta}^\bot(V_b)]\widetilde{\chi}(x,y)
e^{-\frac{2\pi}{a'}z_\mu}
\end{equation}

\noindent with:
$\widetilde{\delta}^\bot(V_b)=\frac{2\pi}{a'}\delta^\bot(V_b)$ and
$\widetilde{\chi}(x,y)=\cos\left[\frac{2\pi}{a'}(x-x_0)\right]+\cos
\left[\frac{2\pi}{a'}(y-y_0) \right]$, a spatial modulation term.
$x_0$ and $y_0$ are the $x$ and $y$ coordinates of the center of
the asperity projected onto the unit cell. Setting $x_0=y_0=0$
locates the asperity and therefore the tip on top of an anion, the
reference ion within the defined unit cell. The above expression
exhibits the expected exponential decaying behavior as a function
of $z_\mu$. The potential is reported in fig.\ref{FIG_POT} for
$a=0.66$~nm, $\delta^\bot=11$~pm and $z_\mu=4$~\AA. The value of
$\delta^\bot$ will be justified in section \ref{SEC_DISCUSSION}.

The expression of
$\sigma_{\mu}=\sigma_{\mu}^{(1)}+\sigma_{\mu}^{(2)}$ can now be
derived from equ.\ref{EQU_SIGMA_MU}. The calculation of
$\sigma_{\mu}^{(1)}$ yields:

\begin{equation}\label{EQU_SIGMA_MU1}
\begin{array}{r}
\sigma_{\mu}^{(1)}=\frac{q}{
a'^2}\cosh[\widetilde{\delta}^\bot(V_b)]\left\{
\widetilde{\chi}(R_a,\theta,\varphi)\mathcal{F}^{(1)}(\theta)-\right.\\
\left.\widetilde{\zeta}(R_a,\theta,\varphi)\mathcal{G}^{(1)}(\theta)\right\}e^{-\frac{2\pi}{a'}(z_\mu+R_a)}\end{array}
\end{equation}

\noindent with $\widetilde{\chi}(r,\theta,\varphi)$, the
expression of $\widetilde{\chi}(x,y)$ in the spherical coordinate
system centered on the asperity and
$\widetilde{\zeta}(r,\theta,\varphi)$, the spherical expression of
another spatial modulation term given by:
$\widetilde{\zeta}(x,y)=(x\sin\left[\frac{2\pi}{a'}(x-x_0)\right]+y\sin
\left[\frac{2\pi}{a'}(y-y_0) \right])/\sqrt{x^2+y^2}$. The
functions $\mathcal{F}^{(1)}$ and $\mathcal{G}^{(1)}$ are written:

\begin{equation}
\mathcal{F}^{(1)}(\theta)=\frac{2a'}{\pi
R_a}\sinh\left(\frac{\eta_\theta}{2}\right)-
8\cos\theta\cosh\left(\frac{\eta_\theta}{2}\right)
\end{equation}

\noindent and:

\begin{equation}
\mathcal{G}^{(1)}(\theta)=8\sin\theta\sinh\left(\frac{\eta_\theta}{2}\right)
\end{equation}

\noindent with: $\eta_\theta= 4\pi R_a\cos\theta/a'$ (see also the
appendix). Thus, it was necessary to assume area (2) as an
infinite plane in order to derive the expression of
$\sigma_{\mu}^{(1)}$ by means of the method of the images. In such
a case, the expression of $\sigma_\mu^{(2)}$ is rigorously derived
from equ.\ref{EQU_SIGMA_MU} and is explicitly given in
ref.[\onlinecite{Note_Nony07e}]. For $x$ and $y$ positions large
compared to $R_a$ however, this description does not fit with the
geometry of the tip apex defined in problem B. The ``infinite
plane" must actually be shrunk down to a spatially limited area
around the asperity, as sketched in fig.\ref{FIG_GEOM}d. This is
made possible when assuming that any planar area is the asymptotic
limit of a sphere with large radius compared to its extension. The
former statement is fulfilled by the condition $R \gg R_a$. The
surface charge density of a sphere with radius $R$ under the
influence of the surface potential $V_s$ is also derivable from
the Kelvin transform. Here, the spherical coordinate system is
centered on the sphere with radius $R$. $\sigma_{\mu}^{(2)}$ is
written:

\begin{equation}\label{EQU_SIGMA_MU2}
\begin{array}{r}
\sigma_{\mu}^{(2)}=\frac{q}{a'^2}\cosh[\widetilde{\delta}^\bot(V_b)]\left\{
\mathcal{F}^{(2)}(\theta)\widetilde{\chi}(R,\theta,\varphi)-\right.\\
\left.\mathcal{G}^{(2)}(\theta)
\widetilde{\zeta}(R,\theta,\varphi)\right\}e^{-\frac{2\pi}{a'}(z_\mu+R_a)}\end{array}
\end{equation}

\noindent The functions $\mathcal{F}^{(2)}$ and
$\mathcal{G}^{(2)}$ are written:

\begin{equation}
\mathcal{F}^{(2)}(\theta)=\left(\frac{a'}{\pi R}-4\cos \theta
\right)e^{-\frac{2\pi}{a'}R(\cos \theta+1)}
\end{equation}

\noindent and:

\begin{equation}
\mathcal{G}^{(2)}(\theta)=4\sin \theta e^{-\frac{2\pi}{a'}R(\cos
\theta+1)}
\end{equation}

\noindent Since $\mathcal{F}^{(2)}(\theta)$ and
$\mathcal{G}^{(2)}(\theta)$ decrease exponentially fast as one
moves away from the foremost position of the tip apex, it can
readily be verified that equ.\ref{EQU_SIGMA_MU2} and the
expression given in ref.[\onlinecite{Note_Nony07e}] do fit for $x$
or $y$ $\in [R_a;3R_a]$. Thus, even in the vicinity of the
asperity, expression \ref{EQU_SIGMA_MU2} can be used instead of
ref.[\onlinecite{Note_Nony07e}]. Therefore, the most part of the
contribution of the sphere to $\sigma_\mu^{(2)}$ is restricted to
a small area that can be assumed as locally planar.

\begin{figure}[t]
 \includegraphics[width=\columnwidth, angle=0]{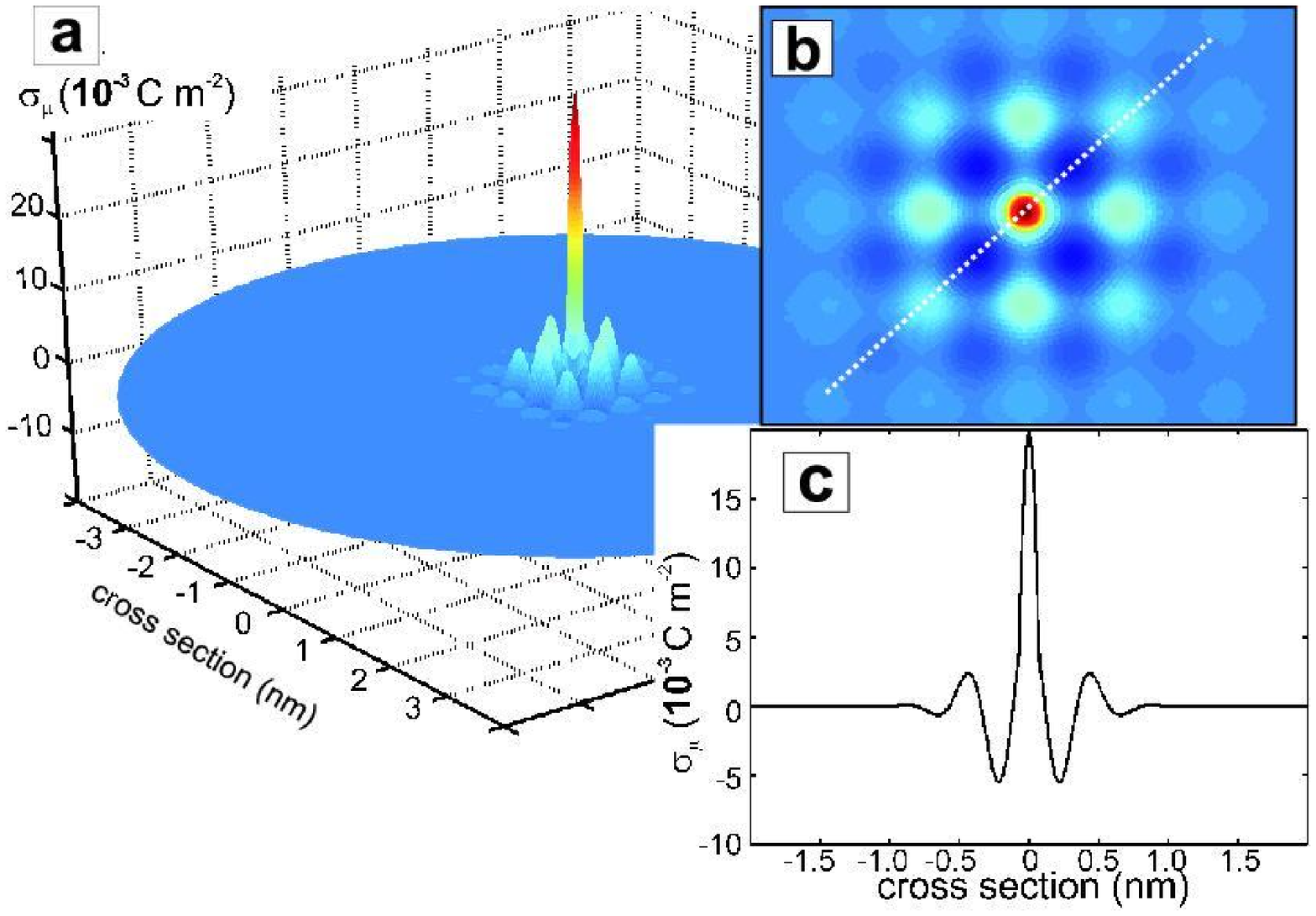}\\
 \caption{(Color online) a- Perspective view of the projection of the microscopic surface charge density $\sigma_\mu$ on the tip calculated on top of an anion from equs.\ref{EQU_SIGMA_MU1} and
 \ref{EQU_SIGMA_MU2} with $R_a=1$~\AA~and $R=5$~nm. $\sigma_\mu$ is strongly increased
 on top of the asperity. Then, owing to the exponential decay of the Madelung surface potential, the surface charge density strongly decreases, which makes the contribution
 of the mesoscopic apex weak. b- Top view of $\sigma_\mu$. The oscillations due to the image charges of the ions at the surface of the crystal are well visible. The most part of the attenuation
 of $\sigma_\mu$ occurs within a single unit cell. Two unit cells apart from the asperity, $\sigma_\mu$ is almost zero. c-. Cross section along the dotted white line shown in b.}\label{FIG_SIGMA_MU}
\end{figure}

The graph of the projection of $\sigma_\mu$ on the tip apex (areas
(1) and (2)) is reported in fig.\ref{FIG_SIGMA_MU} on top of an
anion (positive charge density on the asperity) at a distance
$z_\mu=4$~\AA~from the surface and for $R=5$~nm and $R_a=1$~\AA.
The oscillations of $\sigma_\mu$ at the surface of the tip due to
the image charges of the crystal are readily visible, although
their amplitude decreases exponentially fast along the sides of
the asperity. The relevant oscillations in area (2) spread out in
a circular area with only two unit cells radius. Owing to the
exponential decay of the Madelung surface potential farther away
from the asperity, the tip surface is too far from the crystal
surface to produce relevant image charges. On top of the asperity,
the surface charge density reaches about $2\times
10^{-2}$~C.m$^{-2}$. For a half-sphere with radius $R_a=1$~\AA~and
with $\delta^\bot=11$~pm, this results in a local charge carried
by the tip of about $0.8\times 10^{-2}~q$.

\section{Estimation of the force}\label{SEC_FORCE}
\subsection{Estimation of $F_m$}\label{SEC_FORCE_M}
Owing to the geometry of the problem and since $\sigma_m$ is
equivalent to the surface charge density of an isolated,
conducting sphere, we can now fully evaluate the purely capacitive
component $F_{m}$ to the total force. The force effectively acting
on the mesoscopic part of the tip apex can be derived from the
image charge of the sphere $-4\pi R^2 \sigma_m$ placed at a
symmetric position with respect to the counter-electrode,
\emph{i.e.} at a distance $2h_d$ from the mesoscopic apex:

\begin{equation}
F_{m}=-\frac{(4\pi R^2 \sigma_m)^2}{4\pi\epsilon_0
(2h_d)^2}=-\frac{\pi
R^2}{h_d^2}\widetilde{\epsilon}_d^2\epsilon_0V_b^2
\end{equation}

\noindent With $\epsilon_d \simeq 4.87$
(ref.[\onlinecite{andeen70a}]) and $K=0.9$ (arbitrarily), then
$\widetilde{\epsilon}_d=K\epsilon_d=4.38$. For a typical ac
modulation of the bias of about 0.5~V and considering $R\simeq
5$~nm and $h_d= 5$~mm, the former equation gives an estimation for
the mesoscopic component of the electrostatic force: $F_{m}\simeq
8 \times 10^{-11}$~pN. As suspected, this contribution is
negligible compared to the other two
components\cite{Note_Nony07d}. Therefore the expression of the
total force simplifies to:

\begin{equation}
F=F_{m\mu}+F_{\mu}=\int_\text{tip}{\frac{\sigma_m\sigma_{\mu}}{\epsilon_0}d\mathcal{S}}+\int_\text{tip}{\frac{\sigma_{\mu}^2}{2\epsilon_0}d\mathcal{S}}
\end{equation}

\noindent The reason why the tip has been restricted to its apex
is now clear. Regarding the former capacitive force, owing to the
distance between the tip and the counter-electrode, the
contribution of a macroscopic body in addition to the mesoscopic
apex would not change notably the total force. Regarding the other
force components, $F_{m\mu}$ and $F_{\mu}$, owing to the
exponential decay of the Madelung surface potential, the influence
of the macroscopic body of the tip is expected to be negligible as
well.

\subsection{Estimation of $F_\mu$}\label{SEC_FORCE_MU}
Let us first focus on $F_{\mu}$. The integral must be performed
over the asperity and the mesoscopic sphere around it, areas (1)
and (2), respectively. It is recalled that the two areas do not
overlap each together. For each of them, the spherical coordinate
system must be centered on the corresponding sphere. Thus:

\begin{equation}
\begin{array}{r}
F_{\mu}=F_{\mu}^{(1)}+F_{\mu}^{(2)}=\frac{R_a^2}{2\epsilon_0}\int_{\frac{\pi}{2}}^\pi\int_0^{2\pi}\cos
\theta \sin \theta {\sigma_\mu^{(1)}}^2 d\theta d\varphi+\\
\frac{R^2}{2\epsilon_0}\int_{\frac{\pi}{2}}^{\theta_M}\int_0^{2\pi}\cos
\theta \sin \theta {\sigma_\mu^{(2)}}^2 d\theta
d\varphi\end{array}
\end{equation}

\noindent The radial coordinates are fixed to $r=R_a$ and $r=R$ in
areas (1) and (2), respectively. The polar integration in area (1)
is performed with $\theta \in [\pi/2;\pi]$, whereas in area (2),
it is performed with $\theta \in [\pi/2;\theta_M]$, where
$\theta_M=\pi-\arcsin(R_a/R)$, which ensures the continuity from
(1) to (2). Regarding area (2), the choice of the beginning angle
of the interval ($\pi/2$) does not influence notably the result of
the integration. In other words, owing to the exponential decay of
$\sigma_\mu^{(2)}$, the exact shape of the mesoscopic part of the
tip apex far from the asperity is not relevant. This justifies
\emph{a posteriori} the choice, although simple, of a spherical
geometry. Both integrations over the azimuthal angle must be
performed over $\varphi \in [0; 2\pi]$. Note that in each of the
former integrals, the term $\cos \theta$ stands for the vertical
projection of the force, as stated initially. Although the
integration over the azimuthal angle yield an analytical result,
the integration over $\theta$ does not, which requires to evaluate
some integrals numerically. The expression of $F_{\mu}^{(1)}$ is
written:

\begin{equation}\label{EQU_F_MIC_1}
\begin{array}{r}
F_{\mu}^{(1)}=\frac{q^2R_a^2}{2\epsilon_0
a'^4}\cosh^2[\widetilde{\delta}^\bot(V_b)]e^{-\frac{4\pi}{a'}(z_\mu+R_a)}\left\{
A^{(1)}+\right.\\
\left.B^{(1)}\left[\cos\left(\widetilde{x}_0\right)+\cos\left(\widetilde{y}_0\right)\right]+2C^{(1)}\cos\left(\widetilde{x}_0\right)\cos\left(\widetilde{y}_0\right)\right\}\end{array}
\end{equation}

\noindent with: $\widetilde{x}_0=2\pi x_0/a'$ and
$\widetilde{y}_0=2\pi y_0/a'$, the reduced coordinates of the tip
above the crystal surface. The integral forms of coefficients
$A^{(1)}$, $B^{(1)}$ and $C^{(1)}$ are reported in the appendix as
functions of $R_a$ and $a'$. Taking a typical lattice constant for
alkali halides $a'=a\sqrt{2}/2 \simeq 0.45$~nm and assuming $R_a
\simeq 1$~\AA, we get:

\begin{equation}
A^{(1)} \simeq -130 \hspace{0.5cm} B^{(1)} \simeq -70
\hspace{0.5cm} C^{(1)} \simeq A^{(1)} \simeq -130
\end{equation}

\noindent Thus, $F_\mu^{(1)}$ explicitly depends on the spatial
modulation of the surface potential. Note also the exponential
decay with the distance, actually faster than the distance
dependence of the Madelung surface potential and also the doubling
spatial period term.

Similar integration on area (2) yields:

\begin{equation}
\begin{array}{r}
F_{\mu}^{(2)}=\frac{q^2R^2}{2\epsilon_0
a'^4}\cosh^2[\widetilde{\delta}^\bot(V_b)]e^{-\frac{4\pi}{a'}(z_\mu+R_a)}\left\{
A^{(2)}+\right.\\
\left.B^{(2)}\left[\cos\left(\widetilde{x}_0\right)+\cos\left(\widetilde{y}_0\right)\right]+2C^{(2)}\cos\left(\widetilde{x}_0\right)\cos\left(\widetilde{y}_0\right)\right\}\end{array}
\end{equation}

\noindent The integral forms of coefficients $A^{(2)}$, $B^{(2)}$
and $C^{(2)}$ are derived from those of coefficients $A^{(1)}$,
$B^{(1)}$ and $C^{(1)}$, by replacing $R_a$,
$\mathcal{F}^{(1)}(\theta)$ and $\mathcal{G}^{(1)}(\theta)$ with
$R$, $\mathcal{F}^{(2)}(\theta)$ and $\mathcal{G}^{(2)}(\theta)$,
respectively. The integration is now performed with $\theta \in
[\pi/2;\theta_M]$. With similar parameters than before and setting
$R=5$~nm, we now get:

\begin{equation}
A^{(2)} \simeq -8 \hspace{0.5cm} B^{(2)} \simeq C^{(2)} \simeq 0
\end{equation}

\noindent The striking discrepancy between $A^{(2)}$, $B^{(2)}$
and $C^{(2)}$ is due to the combined contribution of the Bessel
functions and the exponential decay of the functions
$\mathcal{F}^{(2)}(\theta)$ and $\mathcal{G}^{(2)}(\theta)$
occurring in the integral forms of the coefficients $B^{(2)}$ and
$C^{(2)}$. Hence, the force simplifies to:

\begin{equation}\label{EQU_F_MIC_2}
F_{\mu}^{(2)}=\frac{q^2R^2}{2\epsilon_0
a'^4}\cosh^2[\widetilde{\delta}^\bot(V_b)]e^{-\frac{4\pi}{a'}(z_\mu+R_a)}A^{(2)}
\end{equation}

\noindent The spatial modulation of the potential does not
influence the mesoscopic part of tip while scanning the surface.
This contribution acts as a static shift to the total
electrostatic force, similarly as the Van der Waals long-range
interaction for the short-range chemical interactions which are
responsible for the topographic atomic contrast in nc-AFM.

\subsection{Estimation of $F_{m\mu}$}\label{SEC_FORCE_MMU}
The geometrical splitting in terms of areas (1) and (2) used for
the estimation of $F_{\mu}$, can equivalently be applied to
$F_{m\mu}$. Thus:

\begin{equation}
\begin{array}{ll}
F_{m\mu}&=F_{m\mu}^{(1)}+F_{m\mu}^{(2)}=
\\& \frac{R_a^2}{\epsilon_0}\int_{\frac{\pi}{2}}^\pi\int_0^{2\pi}\cos \theta \sin \theta
{\sigma_m^{(1)} \sigma_\mu^{(1)}} d\theta d\varphi+\\ &
\frac{R^2}{\epsilon_0}\int_{\frac{\pi}{2}}^{\theta_M}\int_0^{2\pi}\cos
\theta \sin \theta \sigma_m^{(2)}\sigma_\mu^{(2)} d\theta
d\varphi\end{array}
\end{equation}

\noindent $\sigma_m^{(1)}$ and $\sigma_m^{(2)}$ are the surface
charge densities on areas (1) and (2) within the frame of problem
A. The calculation of $F_{m\mu}^{(1)}$ yields:

\begin{equation}\label{EQU_F_MICMES_1}
\begin{array}{r}
F_{m\mu}^{(1)}=\frac{3\widetilde{\epsilon}_dqR_a^2}{a'^2R}V_b \cosh[\widetilde{\delta}^\bot(V_b)] e^{-\frac{2\pi}{a'}(z_\mu+R_a)}\times \\
D^{(1)}\left[\cos\left(\widetilde{x}_0\right)+\cos\left(\widetilde{y}_0\right)\right]
\end{array}
\end{equation}

\noindent The integral form of $D^{(1)}$ is reported in the
appendix. The numerical integration results in $D^{(1)} \simeq
-15$. Similar integration over area (2) gives a similar expression
of $F_{m\mu}^{(2)}$ by replacing indexes (1) with indexes (2) and
$R_a$ with $R$. The integral form of the coefficient $D^{(2)}$ is
similar to the one of $D^{(1)}$, except that the term $\cos^2
\theta$ is replaced by $\cos \theta$. The numerical integration of
$D^{(2)}$ yields almost zero. This was expected since the
integration is performed over the oscillations of the charge
density $\sigma_\mu^{(2)}$ due to the image charges of the surface
potential (\emph{cf.} fig.\ref{FIG_SIGMA_MU}). On the opposite,
this was not observed in $F_\mu^{(2)}$ because the integration was
performed on the square of $\sigma_\mu^{(2)}$. Therefore
$F_{m\mu}$ is finally written:

\begin{equation}
F_{m\mu}=F_{m\mu}^{(1)}
\end{equation}

\noindent Hence, the former equation states that the coupling of
the force to the bias $V_b$ actually occurs only by means of the
microscopic surface charge density at the foremost part of the
tip, \emph{i.e.} on the asperity. Owing to the integration over
the mesoscopic part of the tip apex and the subsequent
cancellation due to the oscillations of $\sigma_\mu^{(2)}$, no
relevant coupling between the bias voltage and the mesoscopic part
of the tip apex can occur. This aspect strongly suggests that the
effect we are reporting is mainly controlled by the foremost
structure of the tip. The effect is expected to be much enhanced
in case of tips with sharper geometries, particularly those with
apexes including atomically sharp edges.

\begin{figure}[t]
 \includegraphics[width=\columnwidth, angle=0]{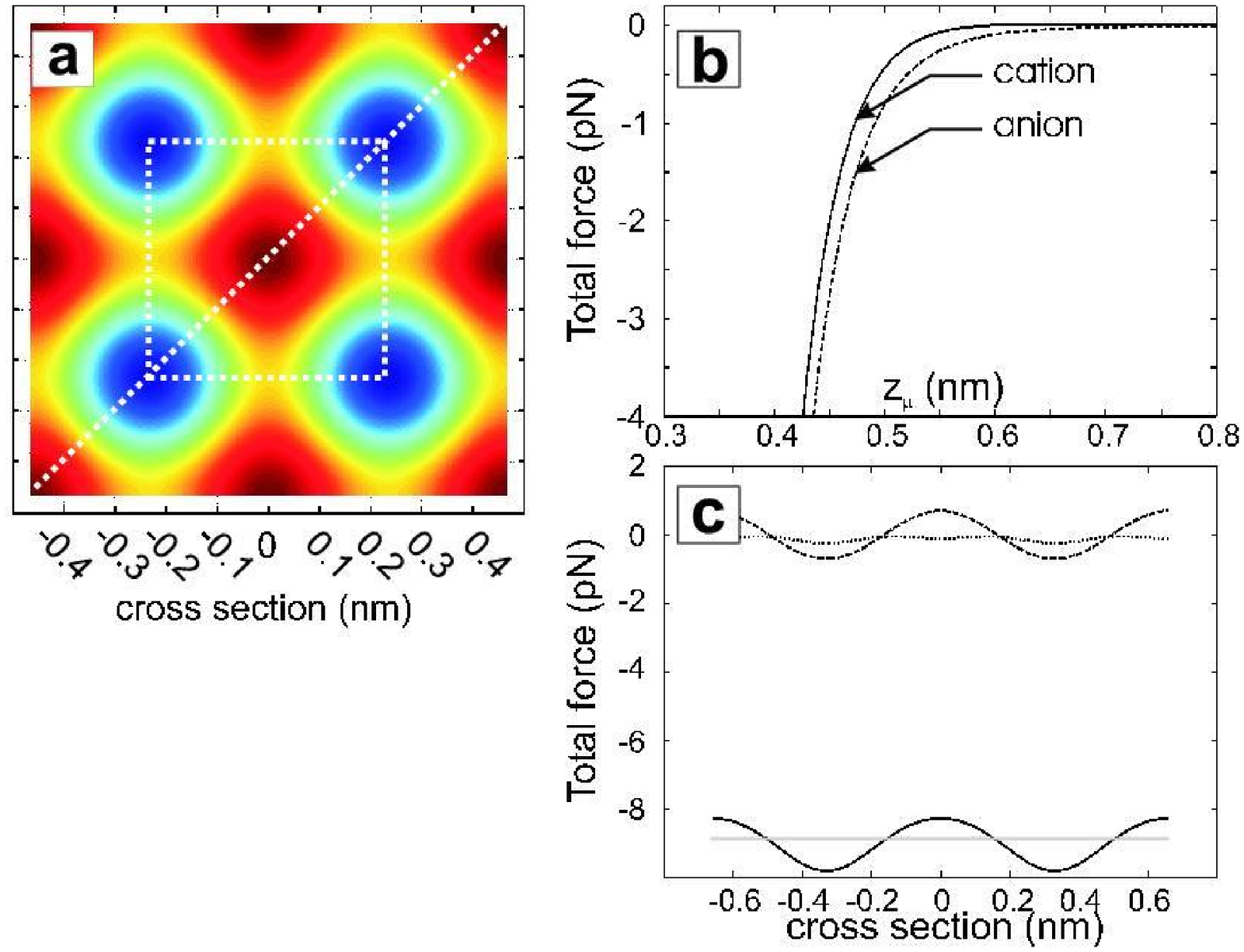}\\
 \caption{(Color online). a- Total electrostatic force over a unit cell calculated from $F=F_{m\mu}^{(1)}+F_{\mu}^{(1)}+F_{\mu}^{(2)}$ for $a=0.66$~nm, $\delta^\bot=11$~pm, $z_\mu=4$~\AA~and $V_b=+1$~V. The unit cell depicted with a dotted line is centered on a cation.
The vertical contrast ranges from -10 (blued spots) to -8~pN
(reddish spots). The force is more repulsive on top of cations
(central ion) than on top of anions, consistently with the bias
polarity. b- $z_\mu$ dependence of the electrostatic force on top
of an anion (dotted curve) and on top of a cation (continuous
curve). c- Cross section along the dotted line shown in a- showing
the total force (thick continuous line), $F_{m\mu}^{(1)}$ (dashed
line), $F_{\mu}^{(1)}$ (dotted line) and $F_{\mu}^{(2)}$ (greyed
line). $F_\mu^{(1)}$ is negligible compared to the
others.}\label{FIG_FORCE}
\end{figure}

The expression for the vertical contribution of the total force
acting on the tip due to the combined influence of the capacitive
coupling and of the Madelung surface potential is finally written:
$F=F_{m\mu}+F_\mu=F_{m\mu}^{(1)}+F_{\mu}^{(1)}+F_{\mu}^{(2)}$,
with $F_{m\mu}^{(1)}$ given by equ.\ref{EQU_F_MICMES_1},
$F_{\mu}^{(1)}$ by equ.\ref{EQU_F_MIC_1} and $F_{\mu}^{(2)}$ by
equ.\ref{EQU_F_MIC_2}. The graph of the total force and of its
components is reported in fig.\ref{FIG_FORCE} for similar
parameters than previously, namely: $a=0.66$~nm,
$\delta^\bot=11$~pm and $z_\mu=4$~\AA. A typical value of bias has
been set, $V_b=+1$~V. Thus, it is visible that the term
$F_\mu^{(1)}$ is negligible compared to others. This is due to the
prefactor $R_a^2/a'^4$ \emph{vs.} $R^2/a'^4$ for $F_\mu^{(2)}$.
The total electrostatic force $F$ finally simplifies to:

\begin{equation}\label{EQU_TOTAL_FORCE}
F=F_{m\mu}^{(1)}+F_{\mu}^{(2)}
\end{equation}

\noindent In fig.\ref{FIG_FORCE}, the force reaches an average
value of about 9~pN (absolute value) and a corrugation of about
2~pN (peak to peak).

\section{IMPLICATIONS FOR KPFM}\label{SEC_DISCUSSION}
\subsection{Estimation of $\delta^\bot(V_b)$}
The net displacement $\delta^\bot(V_b)$ of the topmost ionic layer
 induced by the polarization can be estimated out of the electric field $\overrightarrow{E}$ between the tip and
the surface (\emph{cf.} equ.\ref{EQU_POLARIZATION}). Section
\ref{SEC_PROBLEM_A} has shown that the field induced by the bias
voltage was essentially controlled by the mesoscopic radius of the
tip, $R$, and not by the foremost asperity. Therefore, owing to
equs.\ref{EQU_SIGMA_M2} or \ref{EQU_SIGMA_M1}: $E \simeq V_b/R$.
The expression of $\delta^\bot(V_b)$ can now be deduced using
equ.\ref{EQU_POLARIZATION}:

\begin{equation}\label{EQU_DELTA_PERP}
\delta^\bot(V_b)=\frac{\chi_d}{q}E=V_b\frac{\chi_d}{qR}
\end{equation}

\noindent An order of magnitude for $\delta^\bot(V_b)$ can now be
calculated as follows: with $V_b \simeq 1$~V, $R=5$~nm,
$\alpha_i=70 \times 10^{-30}$~m$^3$ (ref.[\onlinecite{kittel}])
and $n_v=8\sqrt{2}/a^3\simeq 40 \times 10^{27}$~m$^{-3}$ (number
of polarizable ionic species \emph{per} volume unit in a fcc
crystal of KBr with a lattice constant $a=0.66$~nm), we get
$E=2\times 10^8$~V.m$^{-1}$, $\chi_d \simeq 9 \times
10^{-39}$~F.m$^2$ and therefore~$\delta^\bot \simeq 11$~pm.

\subsection{Detected signal in KPFM: connection with the local
CPD}\label{SEC_KPFM} When performing KPFM experiments, the bias
voltage $V_b$ is modulated with a frequency $f_k$ and may as well
include a static component to compensate for the long-range
electrostatic forces. This is the reason why the average value of
the experimental CPD image shown in fig.\ref{FIG_EXP}a reaches
-3.9~V. It is not rare that, on ionic surfaces, many volts are
required to compensate for the long-range electrostatic forces due
to trapped charges while the cleavage of the cristal
\cite{barth06a}. Thus:

\begin{equation}\label{EQU_Vb}
V_b=V_{dc}+V_{ac}\sin(2\pi f_k t)
\end{equation}

\noindent The electrostatic force is thus triggered at $f_k$ and
then detected as an additional low- or high- frequency component
when doing FM- or AM-KPFM, respectively. In both techniques, a
proper dc bias voltage produced by an external controller,
hereafter referred to as $V_{dc}^{(c)}$, is applied between the
tip and the counter-electrode to cancel the modulated component at
$f_k$, \emph{i.e.} the oscillation amplitude of the second bending
eigenmode of the cantilever in AM-KPFM, or the one of the
frequency shift in FM-KPFM. When applied to the tip (\emph{i.e.}
with the counter-electrode grounded), this dc bias is the opposite
of the local CPD defined in equ.\ref{EQU_LOCAL_CPD}:
$V_{dc}^{(c)}=-V_\text{cpd}$.

In order to stick to the AM-KPFM experiments (fig.\ref{FIG_EXP}),
it is necessary to estimate the amplitude of the second eigenmode
of the cantilever and then derive a condition on the dc value of
the bias ultimately able to nullify it. On the one hand, it is
therefore mandatory to check carefully all the occurrences of the
modulated component of the bias voltage in the expression of the
force. This includes explicit dependencies, such as those due to
the polarization, but also implicit ones, as discussed hereafter.
On the other hand, an expression of the oscillation amplitude of
the mode modulated at $f_k$ must be derived.

We first address the problem of explicit and implicit bias
dependencies in the expression of the electrostatic force. Owing
to the explicit $V_b$ dependence in $\delta^\bot$, which has been
kept throughout the description of the model, it can be seen that
the polarization effect is mainly included in the $\cosh$ function
of $F_{m\mu}^{(1)}$ and $F_{\mu}^{(2)}$ through a linear and a
quadratic dependence, respectively (\emph{cf.}
equs.\ref{EQU_F_MICMES_1} and \ref{EQU_F_MIC_2}). Furthermore,
since $\delta^\bot$ is small compared to $a'$,
$\cosh[\widetilde{\delta}^\bot(V_b)]$ can be expanded in series.
To first order:

\begin{equation}
\cosh[\widetilde{\delta}^\bot(V_b)] \simeq 1+
\widetilde{\delta}^\bot(V_b)^2=1+\frac{4\pi^2}{a'^2}\delta^\bot(V_b)^2=1+(\chi'_dV_b)^2
\end{equation}

\noindent with $\chi'_d=2\pi\chi_d/(a'qR)$. Replacing this
expansion in the expressions of the components of the force and
keeping the linear and quadratic terms in $V_b$ yields, with
compact notations:

\begin{equation}
F_{m\mu}^{(1)}=\widetilde{\epsilon}_dK_{m\mu}^{(1)}\Phi_{m\mu}^{(1)}\frac{qV_b}{R}
\end{equation}

\noindent and:

\begin{equation}
F_{\mu}^{(2)}=K_{\mu}^{(2)}\Phi_{\mu}^{(2)}[1+2(\chi'_dV_b)^2]\frac{q^2}{\epsilon_0
a'^2},
\end{equation}

 \noindent where $K_{m\mu}^{(1)}$ and $K_{\mu}^{(2)}$ are two dimensionless coefficients standing for geometrical
 factors of areas (1) and (2), respectively:

\begin{equation}
K_{m\mu}^{(1)}=\frac{3R_a^2}{a'^2}D^{(1)} \hspace{0.5cm}
\text{and} \hspace{0.5cm} K_{\mu}^{(2)}=\frac{R^2}{2a'^2}A^{(2)}
\end{equation}

\noindent $\Phi_{m\mu}^{(1)}$, $\Phi_{\mu}^{(2)}$ are also two
dimensionless coefficients carrying the spatial dependence of each
force component:

\begin{equation}\label{EQU_PHI_MMU1}
\Phi_{m\mu}^{(1)}=e^{-\frac{2\pi}{a'}(z_\mu+R_a)}\left[\cos\left(\widetilde{x}_0\right)+\cos\left(\widetilde{y}_0\right)\right]
\end{equation}

\noindent and:

\begin{equation}
\Phi_{\mu}^{(2)}=e^{-\frac{4\pi}{a'}(z_\mu+R_a)}
\end{equation}

 \noindent Implicit $V_b$ dependencies are now discussed. In the two former equations, particular attention must be paid to $z_\mu$.
 So far, this parameter was defined as the tip-surface distance and set to an
 arbitrary, constant, value. However, when dealing with AM-KPFM, $z_\mu$ is not static but actually coupled to the bias and to the oscillation amplitude of the fundamental eigenmode of the cantilever.
 In the following, for the sake of clarity, the variables related to the fundamental
 eigenmode of the cantilever will be denoted with index ``0" and those
 of the second eigenmode with index ``1".
 Thus, let $z_0$, $z_1$ and $D$ be the instantaneous
 position of the fundamental eigenmode of the cantilever, the instantaneous
 position of the second eigenmode of the cantilever and the distance between the surface and
the equilibrium position of the cantilever at rest, respectively.
Therefore, $z_\mu(t)=D-z_0(t)-z_1(t)$. Hence,
 if $V_b$ has the form given in equ.\ref{EQU_Vb}, one can postulate, to first order: $z_1(t)=A_1\sin(2\pi f_k
 t + \varphi_1)$. $A_1$ and $\varphi_1$ stand for the oscillation amplitude of this mode and its phase lag with respect to the electrostatic actuation, respectively.
 Their exact expressions are not easily derivable, but it must be noticed that $A_1$ must be connected to the amplitude of the modulation, namely $V_{ac}$.
 When $f_k$ accurately matches the actual resonance
 frequency of the second eigenmode\cite{Note_Nony07i}, then
 $\varphi_1=-\pi/2$. $z_0$ is experimentally driven by the
 control electronics of the microscope at the actual resonance frequency of the fundamental mode of the cantilever, $f_0$. It is known that it has an almost harmonic
 behavior of the form: $z_0(t)=A_0\sin(2\pi f_0t-\pi/2)$. Thus, $\Phi_{m\mu}^{(1)}$ and $\Phi_{\mu}^{(2)}$ have a
component at $f_0$ which is further modulated by the dynamics of
the second eigenmode, electrostatically actuated at $f_k$.

We can now propose a self-consistent approximated solution to the
equation of motion for $z_1(t)$ and thus derive the expression of
the oscillation amplitude $A_1$. This equation has the standard
form:

\begin{equation}\label{EQU_DIFF_Z1}
\ddot{ z_1 }(t) +\frac{\omega_1}{Q_1}\dot{z_1}(t) + \omega_1^2
z_1(t)=\frac{F_{ext}}{m_1}+\frac{F_{m\mu}^{(1)}+F_\mu^{(2)}}{m_1},
\end{equation}

\noindent where $F_{ext}$ is an external force oscillating at
$f_0$ which controls the dynamics of $z_0(t)$. Let us assume: i-
$A_1 \ll A_0$, \emph{i.e.} the amplitude of the mode is much
smaller than the one of the fundamental mode, ii- $A_1 \ll a'$,
where $a'=a\sqrt{2}/2$, $a$ being the lattice constant of the
crystal and iii- that the dynamics of $z_1(t)$ is mainly
influenced by components at $f_k$. Assumptions i- and ii- are not
too strong, since the experimental estimations of $A_1$ yield a
few tens of picometers. Assumption i- implies that the dynamics of
the fundamental mode is not much influenced by the one of the
second eigenmode. Hence, the solution of the equation of motion of
$z_0(t)$ has indeed the form postulated above. Assumption ii-
allows us to linearize $\Phi_{m\mu}^{(1)}$ and $\Phi_{\mu}^{(2)}$
with respect to $z_1(t)$. Finally, assumption iii-, which is
consistent with the postulated solution for $z_1(t)$,
$z_1(t)=A_1\sin(2\pi f_k t+\varphi_1)$, simplifies further the
above equation of motion. Now, owing to the former assumptions,
equ.\ref{EQU_DIFF_Z1} can be solved by injecting the postulated
expressions of both eigenmodes and keeping only the terms
oscillating at $f_k$. For that purpose, the exponential term
wherein $z_0(t)$ occurs must be expanded in Fourier series. Then,
the only possibility to preserve terms at $f_k$ is to keep the
lone static component of the Fourier expansion, hereafter referred
to as $a_0$ (expansion of $\Phi_{m\mu}^{(1)}$) and $b_0$ (exansion
of $\Phi_\mu^{(2)}$). After linearization, $\Phi_{m\mu}^{(1)}$ and
$\Phi_{\mu}^{(2)}$ can finally be written as:

\begin{equation}\label{EQU_TOT_PHIMMU1}
\begin{array}{r}
\Phi_{m\mu}^{(1)}\overbrace{\simeq}^\text{i-, ii-}
\left[1-\frac{2\pi}{a'}z_1(t)
\right]e^{-\frac{2\pi}{a'}(D-A_0+R_a)}\times
\\
\{a_0+\underbrace{\sum_{n=1}^{\infty}a_n\cos(2\pi n f_0
t)}_{\text{neglected, owing to
iii-}}\}\left[\cos\left(\widetilde{x}_0\right)+\cos\left(\widetilde{y}_0\right)\right]
\end{array}
\end{equation}

\noindent and:

\begin{equation}\label{EQU_TOT_PHIMU2}
\Phi_\mu^{(2)}\simeq \left[1-\frac{4\pi}{a'}z_1(t)
\right]e^{-\frac{4\pi}{a'}(D-A_0+R_a)}
\{b_0+\underbrace{\sum_{n=1}^{\infty}b_n\cos(2\pi n f_0
t)}_{\text{neglected, owing to iii-}}\}
\end{equation}

\noindent With:

\begin{equation}
\begin{array}{l}
a_n=e^{-\frac{2\pi}{a'}A_0}I\left[n,{\frac{2\pi}{a'}A_0}\right]\\
b_n=e^{-\frac{4\pi}{a'}A_0}I\left[n,{\frac{4\pi}{a'}A_0}\right]
\end{array}
\end{equation}

\noindent $I$ is the modified Bessel function of the first kind.
When replacing equs.\ref{EQU_TOT_PHIMMU1} and \ref{EQU_TOT_PHIMU2}
with $z_1(t)=A_1\sin(2\pi f_kt+\varphi_1)$ in
equ.\ref{EQU_DIFF_Z1}, it is possible to derive an expression for
$A_1$. The condition on $V_{dc}^{(c)}$ to match $A_1=0$ is finally
written:

\begin{equation}\label{EQU_COMPENSATED_CPD}
V_{dc}^{(c)}=-\frac{{\widetilde{\epsilon}}_d \epsilon_0 a'^2 }{4R
q
{\chi'_d}^2}\frac{a_0}{b_0}\frac{K_{m\mu}^{(1)}}{K_\mu^{(2)}}e^{\frac{2\pi}{a'}
(D+R_a-A_0)}[\cos(\widetilde{x_0})+\cos(\widetilde{y_0})]
\end{equation}

\begin{figure}[t]
 \includegraphics[width=\columnwidth, angle=0]{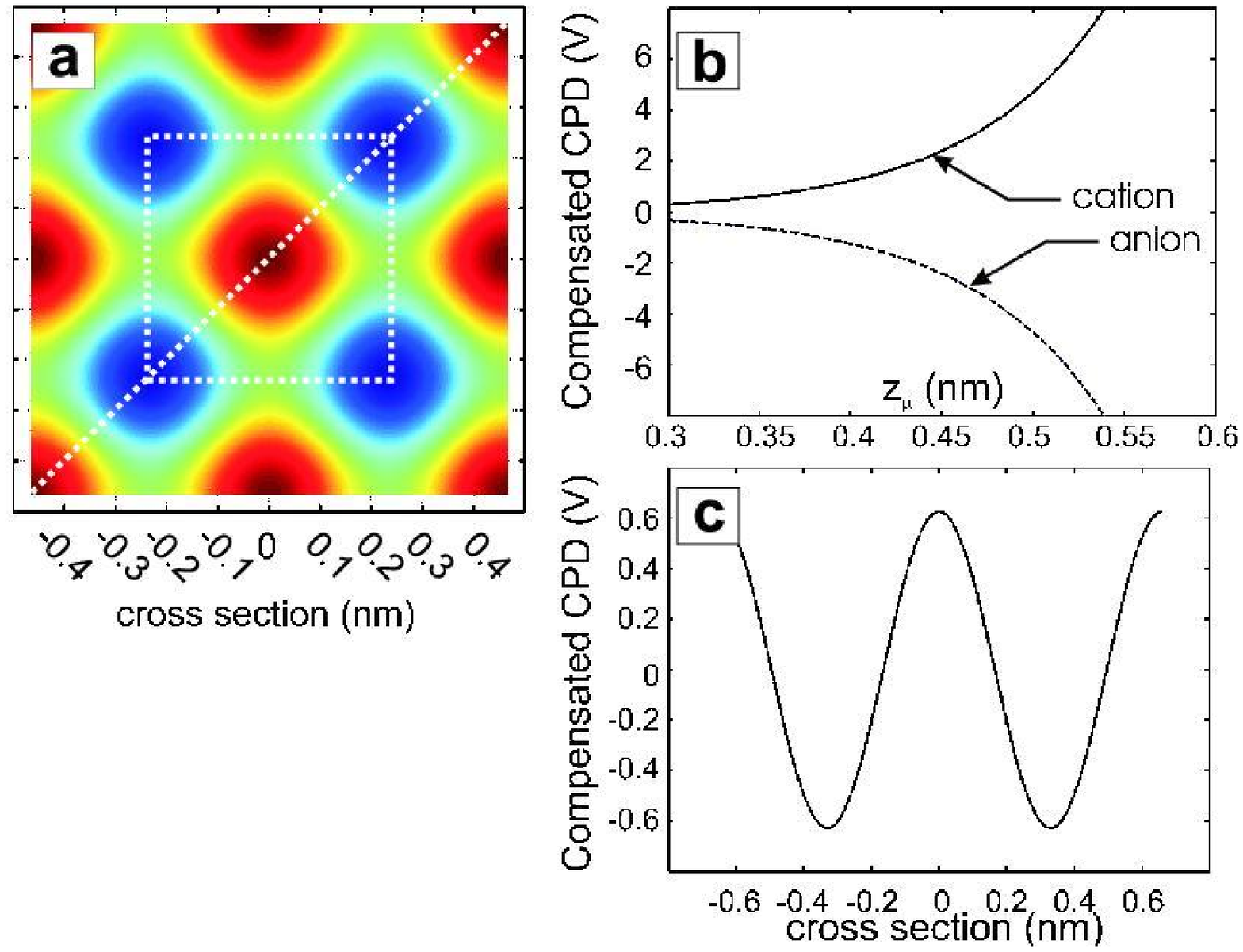}\\
 \caption{(Color online). a- $V_{dc}^{(c)}$ bias voltage required to compensate the local CPD calculated from equ.\ref{EQU_COMPENSATED_CPD} for $a=0.66$~nm,
 $\delta^{\bot}=11$~pm, $A_0=5$~nm and $z_\mu=3.5$~\AA. The vertical contrast ranges from -0.6 (blued spots) to +0.6~V (reddish spots). The unit cell depicted
 with a dotted line is centered on a cation. b- Distance dependence of the potential on top of an anion (dashed curve) and
 on top of a cation (continuous curve). c- Cross section along the dotted line in a-.}\label{FIG_COMPENSATED_CPD}
\end{figure}

\noindent The graph of $V_{dc}^{(c)}$ is reported in
fig.\ref{FIG_COMPENSATED_CPD} for $a=0.66$~nm,
$\delta^\bot=11$~pm, $A_0=5$~nm (hence $a_0=0.0487$ and
$b_0=0.0344$) and $(D-A_0)=3.5$~\AA. The value of $A_0$ has been
chosen consistently with the experimental conditions. On the
contrary, the tip-surface distance has been chosen arbitrary, but
however in a range where the atomic contrast is usually
experimentally achieved. In fig.\ref{FIG_COMPENSATED_CPD}, the
lateral periodicity of the underlying lattice is readily visible,
but surprisingly, the potential scales between -0.6 to +0.6~V from
an anionic to a cationic site, respectively. At similar height,
this is three times larger than the Madelung surface potential
(\emph{cf.} fig.\ref{FIG_POT}b). The comparison with the
experimental results is more severe since the theoretical
prediction is one order of magnitude larger. At this point, it is
recalled that the strong tip geometry dependence of the problem
makes a straightforward comparison between the theoretical
prediction (equ.\ref{EQU_COMPENSATED_CPD}) and the experimental
results difficult, since our analytical expression of
$V_{dc}^{(c)}=-V_{cpd}$ relies on a somewhat unrealistic tip.

\subsection{Experimental implications for AM-KPFM}

The figure and the above formula show that the bias voltage to be
applied on the tip to compensate for the electrostatic force is
governed by three main factors:
\begin{itemize}
    \item the dielectric properties of the sample such as its
dielectric permittivity and lattice constant.
    \item a subtle balance
between mesoscopic and microscopic geometric factors of the tip.
    \item a lateral periodicity similar to the
Madelung surface potential of the crystal.
\end{itemize}

A straightforward consequence is that the atomic corrugation of
the CPD reported experimentally might stand for the spatial
fluctuations of the Madelung surface potential, however with an
amplitude that depends on the surface polarization and hence on
the applied ac voltage. It is also important to notice that the
CPD compensation is proportional to $K_{m\mu}^{(1)}$, \emph{i.e.}
to the asperity size. The former being also the source of the
coupling between the tip/dieletric/counter-electrode capacitor and
the Madelung surface potential, \emph{i.e.} the source of the KPFM
signal, the atomic contrast of the CPD is therefore closely
connected with the geometry of the very foremost part of the tip.
This is consistent with the short-range character of the
interaction. But, on the other hand, the explicit dependence with
geometric factors of the tip, unambiguously proves that
quantitative measurements of the local CPD are unlikely to be
performed in KPFM, unless the tip geometry be accurately known,
which is practically never true.

Furthermore, although equ.\ref{EQU_COMPENSATED_CPD} explicitly
exhibit a distance dependence, consistently with the experimental
observations in the above mentioned references, an increase of the
compensated CPD as a function of the distance is nevertheless
surprising. The residual exponential dependence originates from
$\Phi_{\mu}^{(2)}$, \emph{i.e.} from the influence of the Madelung
surface potential on the mesoscopic tip apex. As discussed above,
the CPD compensation being partly governed by the asperity, a
tip-surface distance increase $\Delta z_\mu$ produces a decrease
of the related force $F_{m\mu}^{(1)}$ proportional to $\exp{(-2\pi
\Delta z_\mu/a')}$ (\emph{cf.} $\Phi_{m\mu}^{(1)}$,
equs.\ref{EQU_PHI_MMU1} or \ref{EQU_TOT_PHIMMU1}). This abrupt
change is compensated by an equivalent exponential increase of the
compensated CPD. This process can obviously not occur at any
tip-surface distance. Prior to being cancelled, the $f_k$
component must stand for a measurable signal. Therefore the above
discussion stands within a narrow range of distances from the
surface, typically a few times the asperity radius.

Nevertheless, an increase of the measured local CPD as a function
of the distance is still expected to occur if the distance
dependence in $\Phi_\mu^{(2)}$ decays faster than the one in
$\Phi_{m\mu}^{(1)}$. In other words $V_{dc}^{(c)}$ must increase
with the distance as soon as the distance dependence of the force
induced by the influence of the Madelung surface potential on the
mesoscopic part of the tip decays faster than the one of the force
induced on the asperity due to the capacitive coupling with the
image charges of the surface.

Let us finally point out that such a distance dependence of the
CPD might make the experimental achievement of the atomic contrast
easier, which strengthens the argument of an intrinsic imaging
process of the local CPD. Indeed, no major tip and/or surface
distortion is expected to occur in an equivalent range of distance
($\geq 3$~\AA, ref.[\onlinecite{Note_Nony07h}]). In that case,
instabilities due to adsorbed and/or mobile atomic or ionic
species at the tip apex are less likely to occur, which makes the
imaging process robust, as experimentally observed.

To conclude, the analytical approach, although restricted to a tip
with a basic geometry, remains helpful, primarily because it
provides an expression of the short-range electrostatic force that
can be connected to the nc-AFM-KPFM simulator. There are obvious
limitations to our approach, the most important one being the use
of classical, continuous electrostatics to treat the angstrom-size
nanoasperity. This obviously must break down at a certain point,
and be replaced by a proper quantum mechanical treatment of the
problem. In the near future, the electrostatic model should be
extended to a bit more complex systems such as local dipoles,
charges or defects at the surface and at steps of ionic crystals.

\section{CONCLUSION}
The aim of this work was to provide a consistent approach to
describe the short-range electrostatic force between the tip of an
nc-AFM and the (001) surface of a perfect ionic crystal. In order
to develop an analytical expression for the total electrostatic
force, the tip has been restricted to a simple geometry and the
influence of the sample has been described by means of its
Madelung surface potential. In such a way, an analytical solution
for the total electrostatic force was found within the
boundary-value problem assuming a thick dielectric sample and an
infinite top-layer of ionic surface charges.

Two major contributions to the electrostatic force can be
extracted: the first stands for a coupling term between the
microscopic structure of the tip apex and the capacitor formed
between the tip, the dielectric ionic crystal and the
counter-electrode due to the bias voltage $V_b$; the second term
depicts the influence of the fluctuations of the Madelung surface
potential arising at the surface of the ionic crystal on the
mesoscopic part of the tip, independently from its microscopic
structure. The former has the lateral periodicity of the Madelung
surface potential whereas the latter only acts as a static
component which shifts the total force.

Beyond the dielectric properties of the crystal, which are
explicitly included in the model, the ionic polarization of the
sample due to the influence of the bias voltage applied to the
tip/counter-electrode capacitor is mainly responsible for the
atomic contrast of the KPFM signal. Typical orders of magnitude
give a net displacement of the ions of about $\pm$10~pm from their
equilibrium positions. Note that this displacement only occurs if
a tip-sample bias is applied (ac or dc), which is always the case
in KPFM experiments.

A detailed analysis of the bias voltage required to compensate for
the electrostatic force shows that the compensated CPD has the
lateral periodicity of the Madelung surface potential. However,
there is a strong dependence on the tip geometry, the applied
modulation voltage as well as the tip-sample distance, which can
even lead to an overestimation of the real surface potential.

For a quantitative evaluation of KPFM results, it is thus
essential to account for all the parameters of the experiment,
among which the tip shape. The analytical expression developed in
this work provides an alternative tool to elucidate the contrast
formation in KPFM on ionic crystals, and in combination with the
nc-AFM simulator it might enable us to interpret our results more
accurately.

\section*{APPENDIX}
After integration of $F_{\mu}^{(1)}$ over the azimuthal angle
$\varphi \in[0;2\pi]$ , we have:

\begin{equation}
\begin{array}{r}
F_{\mu}^{(1)}=\frac{q^2R_a^2}{2\epsilon_0
a'^4}\cosh^2[\widetilde{\delta}^\bot(V_b)]e^{-\frac{4\pi}{a'}(z_\mu+R_a)}\left\{
A^{(1)}+\right.\\
\left.B^{(1)}\left[\cos\left(\widetilde{x}_0\right)+\cos\left(\widetilde{y}_0\right)\right]+2C^{(1)}\cos\left(\widetilde{x}_0\right)\cos\left(\widetilde{y}_0\right)\right\}\end{array}
\end{equation}

\noindent where:

\begin{equation}
A^{(1)}= \pi\int_{\frac{\pi}{2}}^{\pi}\cos\theta \sin \theta
[2\mathcal{F}^{{(1)}^2}(\theta)+\mathcal{G}^{{(1)}^2}(\theta)]d\theta
\end{equation}

\begin{equation}
\begin{array}{r}
B^{(1)}=\pi\int_{\frac{\pi}{2}}^{\pi}\cos\theta \sin \theta
\left\{\left(\mathcal{F}^{{(1)}^2}(\theta)-\frac{\mathcal{G}^{{(1)}^2}(\theta)}{2}\right)J_0\left(\eta_\theta \right) -\right.\\
\left.2\mathcal{F}^{(1)}(\theta)\mathcal{G}^{(1)}(\theta)J_1\left(\eta_\theta
\right)+\frac{\mathcal{G}^{{(1)}^2}(\theta)}{2}J_2\left(\eta_\theta
\right) \right\}d\theta\end{array}
\end{equation}

\noindent and:

\begin{equation}
\begin{array}{r}
C^{(1)}=\pi\int_{\frac{\pi}{2}}^{\pi}\cos\theta \sin \theta
\left\{2\mathcal{F}^{{(1)}^2}(\theta)J_0\left(\eta'_\theta \right) -\right.\\
\left.2\sqrt{2}\mathcal{F}^{(1)}(\theta)\mathcal{G}^{(1)}(\theta)J_1\left(\eta'_\theta \right)+\right.\\
\left.\mathcal{G}^{{(1)}^2}(\theta)J_2\left(\eta'_\theta \right)
\right\}d\theta\end{array}
\end{equation}

\noindent with: $\eta_\theta=4\pi R_a\sin \theta /a'$ and
$\eta'_\theta=2\sqrt{2}\pi R_a\sin \theta /a'$. $J_0$, $J_1$ and
$J_2$ are the Bessel functions of the first kind.

Regarding $F_{m\mu}^{(1)}$, the integration over the azimuthal
angle $\varphi \in[0;2\pi]$ yields:

\begin{equation}
\begin{array}{r}
F_{m\mu}^{(1)}=\frac{3\widetilde{\epsilon}_dqR_a^2}{a'^2R}\cosh[\delta_\bot(V_b)]V_b D^{(1)}\left[\cos\left(\frac{2\pi}{a'}x_0\right)+\right.\\
\left.\cos\left(\frac{2\pi}{a'}y_0\right)\right]
\end{array}
\end{equation}

\noindent where:

\begin{equation}
\begin{array}{r}
D^{(1)}=-2\pi\int_{\frac{\pi}{2}}^\pi \cos^2 \theta \sin \theta
\left\{\mathcal{F}^{(1)}(\theta)J_0\left(\frac{\eta_\theta}{2}\right)-\right.\\
\left.\mathcal{G}^{(1)}(\theta)J_1\left(\frac{\eta_\theta}{2}\right)
\right\} d\theta \end{array}
\end{equation}

\section*{References}
\bibliographystyle{prbsty}

\begin{thebibliography}{10}
\bibitem{Note_Nony07b}
The ionic polarizability $\alpha_i$ is connected to the dielectric
permittivity
  $\epsilon_d$ by Claussius-Mossotti's equation.
\bibitem{Note_Nony07f}
The latter is obtained by anti-symmetrical transform of the
charges layer with
  respect to the counter-electrode symmetry plane.
\bibitem{Note_Nony07c}
This is also true because it has been assumed that the ionic
crystal carried no
  net charge. Therefore there is no long-range electrostatic interaction.
\bibitem{Note_Nony07g}
The latter is now obtained by anti-symmetrical transform of the
layer of ionic
  charges into the asperity.
\bibitem{Note_Nony07e}
The expression of $\sigma_\mu^{(2)}(r>R_a,\theta=\pi/2, \varphi )$
for an
  infinite planar geometry is:
  $\sigma_\mu^{(2)}=\frac{q}{a'^2}\cosh[\delta^\bot(V_b)]\{4\widetilde{\chi}(R%
,\pi/2,\varphi)-4\frac{R^3}{r^3}\widetilde{\chi}(R^2/r,\pi/2,\varphi)\}$.
\bibitem{Note_Nony07d}
As a matter of fact, $h_d$ being very large, it can readily be
verified that
  this result is always true whatever is the exact shape of the mesoscopic part
  of the tip apex.
\bibitem{Note_Nony07h}
In ref.[\onlinecite{hoffmann07a}], the foremost unstable atom of
the tip which
  is responsible for the contrast in dissipation is triggerred at a distance of
  about 3~\AA~from the surface. Above that threshold, no instability is
  observed.
\end{thebibliography}

\end{document}